\documentclass[aps,pre,twocolumn,showpacs,showkeys,groupedaddress,floatfix]{revtex4-1}
\usepackage[T2A]{fontenc}
\usepackage[cp1251]{inputenc}
\usepackage[english]{babel}
\usepackage{graphicx}
\usepackage{subfigure}
\usepackage{multirow}
\usepackage{amsmath}
\usepackage{amssymb}

\graphicspath{{fig/}}

\begin{document}
\title{Percolation of linear $k$-mers on square lattice: from isotropic through partially ordered  to completely  aligned state}

\author{Yuri Yu. Tarasevich}
\email[Corresponding author: ]{tarasevich@aspu.ru}
\affiliation{Astrakhan State University, 20a Tatishchev Street, 414056
Astrakhan, Russia}

\author{Nikolai I. Lebovka}
\email{lebovka@gmail.com}
\affiliation{Institute of
Biocolloidal Chemistry named after F.D. Ovcharenko, NAS of Ukraine,
42, Boulevard Vernadskogo, 03142 Kiev, Ukraine}

\author{Valeri V. Laptev}
\email{serpentvv@mail.ru} \affiliation{Astrakhan State Technical
University, 16 Tatishchev Street, 414025 Astrakhan, Russia}
\affiliation{Astrakhan State University, 20a Tatishchev Street,
414056 Astrakhan, Russia}

\date{\today}

\begin{abstract}
Numerical simulations by means of Monte Carlo method and finite-size scaling analysis have been performed to study the percolation behavior of linear $k$-mers (also denoted in the literature as rigid rods, needles, sticks) on two-dimensional square lattices $L \times L$ with periodic boundary conditions. Percolation phenomena are investigated for anisotropic relaxation random sequential adsorption of linear $k$-mers. Especially, effect of anisotropic placement of the objects on the percolation threshold has been investigated.  Moreover, the behavior of percolation  probability $R_L(p)$ that a lattice of size $L$ percolates at concentration $p$ has been studied in details in dependence on $k$, anisotropy and lattice size $L$. A nonmonotonic size dependence for the percolation threshold has been confirmed in isotropic case. We propose a fitting formula for percolation threshold $p_c = a/k^{\alpha}+b\log_{10} k+ c$, where $a$, $b$, $c$, $\alpha$ are the fitting parameters varying with anisotropy. We predict that for large $k$-mers  ($k\gtrapprox 1.2\times10^4$)  isotropic placed at the lattice,  percolation cannot occur even at jamming concentration.
\end{abstract}


\maketitle

\section{\label{sec:introduction}Introduction}
Percolation deals with the properties of disordered media.Such media can be composed of the objects placed in a space. The objects can connect with each other and form clusters. If object concentration is large enough, infinitely large cluster occurs. Such a concentration is known as a percolation threshold. The properties of media are considerably different below and above percolation threshold. If objects are placed in a space purely randomly, the percolation is called random or Bernoulli percolation. Moreover, different correlations or constrains may be applied to the space distribution of the objects. The media composed in such a way may be partially disordered and anisotropic. Very often, a discrete space (lattice) is utilized to simplify consideration. In this case, the cluster-forming objects are sites of the lattice. Percolation of the point objects (singly occupied site) on different lattices in plane and multidimensional space is more intensively studied. Percolation of the objects occupying several nearest sites is studied much worse. The examples of such objects are linear, cyclic and branched $k$-mers, i.e. $k$ nearest sites. The huge amount of publications is devoted to both theoretical and applied aspects of percolation (see, e.g.,~\cite{Stauffer,Sahimi,Grimmett}). During the past few decades, percolation of the anisotropic penetrable and impenetrable objects (rods, sticks, linear $k$-mers, ellipsoids etc.) has been intensively investigated. In our overview, we restrict ourselves to the works devoted to the percolation of linear objects on a lattice.

Mainly, the studies are devoted to the isotropic problem on a square lattice when the $k$-mers  with horizontal and vertical orientations are  deposited with equal probability. A computer-simulation model for linear $k$-mers ($k=1 \dots 20$) showed that percolation threshold $p_c$ decrease with increasing of the chain-length $k$ as $1/k^{0.5}$~\cite{Becklehimer1992}. The percolation exponents  (order parameter, susceptibility, and correlation length exponents) seem  to remain unchanged.

The study of the percolative properties of systems generated by a random sequential adsorption (RSA) of $k$-mers ($k=1 \dots 40$) have been performed by Leroyer and Pommiers~\cite{Leroyer1994}. They demonstrated that as the segment length grows, the percolation threshold $p_c$ decreases, goes through a minimum and then increases slowly for large $k$ ($k\geq 16$).

Later on, Kondrat and P\c{e}kalski~\cite{KondratPre63} extended the studies percolation and jamming of the same problem to the $k$-mer length in the interval $k=1 \dots 2000$. The authors showed that the jamming threshold decreases monotonically  approaching the asymptotic value of $p_j=0.66\pm 0.01$ at large $k$ and percolation threshold $p_c$ is a nonmonotonic function of the length $k$, showing a minimum for a certain length of the $k$-mers ($k=13$). However, these results for very large needles cannot be treated as accurate because of moderate size of the studied lattices ($L\leq 2500$) and possibility of large finite-size corrections.

The details of the monotonic behavior of the percolation threshold for small $k$-mer length ($k\leq 15$) have been widely discussed in the literature~\cite{Vandewalle2000, Cornette2003,Cornette2006, Cornette2006a}. Percolation and jamming phenomena have been investigated for $k$-mer length within the interval $k=1 \dots 10$ by Vandewalle et al.~\cite{Vandewalle2000}. The authors conjectured presence of some universal connection in the geometry of jamming and percolation that resulted in constancy of the ratio of percolation and jamming concentration $p_c/p_j$ ($\simeq 0.62$) for all sizes of $k$-mers. The following equation for the percolation threshold as a function of $k$-mer length has been proposed
\begin{equation}\label{eq:Vandewalle}
p_c = C \left (1 - \gamma \left ( \frac{k - 1}{k} \right)^2 \right ),
\end{equation}
where $C$ and $\gamma$ are the constants.

Cornette et al.~\cite{Cornette2003} performed the finite-size scaling tests and shown that the $k$-mer problem in all the studied cases, belongs to the random percolation universality class. They fitted the data for the $k$-mers ($k=1 \dots 15$) with following exponential equation:
\begin{equation}\label{eq:RP}
p_c = p_c^\infty + \Omega \exp \left(-\frac{k}{\kappa} \right ),
\end{equation}
where $p_c^\infty= 0.461 \pm 0.001$, $\Omega = 0.197 \pm 0.02$,  and $\kappa = 2.775 \pm 0.02$ are the fitting parameters. $p_c^\infty$ is the expected value in the limit $k \to \infty$.

Recently, these problems have been extended for partially ordered $k$-mer (when the particles with horizontal and vertical orientations can be deposited with of unequal probability)~\cite{Cherkasova,Lebovka2011PRE,Longone2012}. The effects dimer alignment on
percolation and jamming phenomena on a square lattice has been investigated by Cherkasova et al.~\cite{Cherkasova}. The influence of dimer alignment on the electrical conductivity has been examined, too. The effect of $k$-mer alignment on the jamming threshold has been extensively examined for the $k$ in  the interval $1 \dots 256$~\cite{Lebovka2011PRE}. The percolation behavior for the $k$-mer length in the interval $k=1 \dots 15$ has been studied recently by Longone et al.~\cite{Longone2012}. Only two particular cases have been studied in the work, i.e. the isotropic case and the completely ordered case (all $k$-mers  are aligned along the given direction). In both cases, the percolation threshold is monotonic decreasing function of the $k$-mer length $k$.

The numbers of a numerical studies have been recently devoted to the analysis of equilibrium properties in systems of $k$-mers~\cite{Ghosh2007,Matoz-Fernandez2008,Matoz-Fernandez2008a,Matoz-Fernandez2011}. The equilibrium systems have been simulated using the deposition-evaporation dynamics. The studies showed existence of a orientationally ordered phase (nematic phase)  for long $k$-mers. The universality class for the percolation and isotropic-nematic phase transition have been found to be the same as of the random percolation and Ising models.  The non-monotonic size dependence has been observed for the percolation threshold of unaligned $k$-mers, it goes through a minimum at $k\simeq 5$, and asymptotically converges towards a definite value $p_c\simeq 0.54$ for large fully aligned $k$-mers~\cite{Matoz-Fernandez2012}. It has been interpreted as a consequence of the isotropic-nematic phase transition occurring in the system for large values of $k$.

Except pure theoretical interest, such considerations may have different applications. For instance,  the percolation approach is suitable to describe physical and chemical properties of  monolayers formed during adsorption of the polymer chains~\cite{Zerko2012}.
Another possible application is connected with the nanotechnologies (see, e.g.,~\cite{Kyrylyuk2008, Kyrylyuk2008correction}). Recently, the current progress on the production of aligned single-walled carbon nanotubes (SWCNTs) has been reviewed by Ma et al.~\cite{Ma2011}. The semiempirical theories of composites containing randomly oriented anisotropic inclusions (needle, prolate or oblate spheroid, sphere, or disk)
have been developed  and they are useful for prediction of effective electrical or thermal conductivities of multi-walled carbon nanotube composites~\cite{Gao2007, Kyrylyuk2008,*Kyrylyuk2008correction, Pan2011, Vovchenko2011, Kyrylyuk2011}. The first experiments evidenced the lowering of the threshold respective to isotropic systems~\cite{Carmona1980}. The experiments for random stick patterns obtained by photolithographic techniques supported the universality hypothesis for 2d systems~\cite{Noh1986}. The universality concept has been also confirmed in experiments with the aluminum film containing the insulating ellipsoids with the same direction of the major axis~\cite{Han1991}.

This work discusses the percolation behaviour of linear $k$-mers on square lattice with different degree of alignment characterized by order parameter. We try to clear the uncertainty in question about the presence or absence the nonmonotonic $k$-dependence for the percolation threshold by studying the systems with $k$ varies from 1 up to 512.

In our work, we try to find the answers to the  questions listed below
\begin{enumerate}
\item Are the Eqs.~\ref{eq:Vandewalle} and \ref{eq:RP} valid for the very long linear objects or they work only for rather short objects?
\item How does anisotropic placement of the objects effect the percolation threshold?
\end{enumerate}

The rest of paper is arranged as follow. In Section~\ref{sec:model}, we describe our model and the details of simulation. The obtained results are discussed in Section~\ref{sec:results}. We summarize the results and conclude our paper in Section~\ref{sec:concl}.

\section{\label{sec:model}Description of models and details of simulations }

The problem of linear $k$-mers of length $2^n$, where $n = 1,2,\dots ,9$, on the square lattices of size $L \times L$ has been studied. Linear lattice size, $L$, varies from 100 to 19200 in different simulations. Periodic boundary conditions in vertical and horizontal directions have been applied, i.e.  percolation on a torus has been considered.

\subsection{\label{sec:model2}Filling of the lattice by $k$-mers}
The relaxation random sequential adsorption (RRSA) model~\cite{Lebovka2011PRE} has been used to place the $k$-mers on a lattice. In this model, there is an infinitely large reservoir filled with $k$-mers oriented with given and fixed anisotropy. The $k$-mer is taken from the reservoir and an attempt of its deposition is carried out starting from a lattice site selected at random until the object is deposited. In contrast with the conventional random sequential adsorption (RSA) model, when any unsuccessful attempt is rejected and other object is selected for deposition, RRSA model ensures that anisotropy of the deposit is the same as the anisotropy of the objects suspended in the reservoir~\cite{Lebovka2011PRE}.

The degree of anisotropy is characterized by the order parameter $s$ defined as
\begin{equation}\label{eq:S}
    s =  \left|\frac{N_| - N_-}{N_| + N_-}\right|,
\end{equation}
where $N_|$ and $N_-$ are the numbers of $k$-mers oriented in vertical and horizontal directions, respectively.

For isotropic system, $s=0$, the numbers of vertical and horizontal $k$-mers are the same, and for totally aligned system, $s=1$, all $k$-mers are aligned in vertical direction.
For these two marginal cases, RRSA and RSA models are absolutely identical~\cite{Lebovka2011PRE}.

The Mersenne twister random number generator~\cite{Matsumoto} with a period of $2^{19937} - 1$ has been exploited to generate positions and orientations of the deposited objects.

\subsection{\label{sec:model1}Determination of percolation threshold}

\begin{figure}[!htbp]
\centering
\subfigure[Crossing clusters]{\includegraphics*[width=0.9\linewidth]{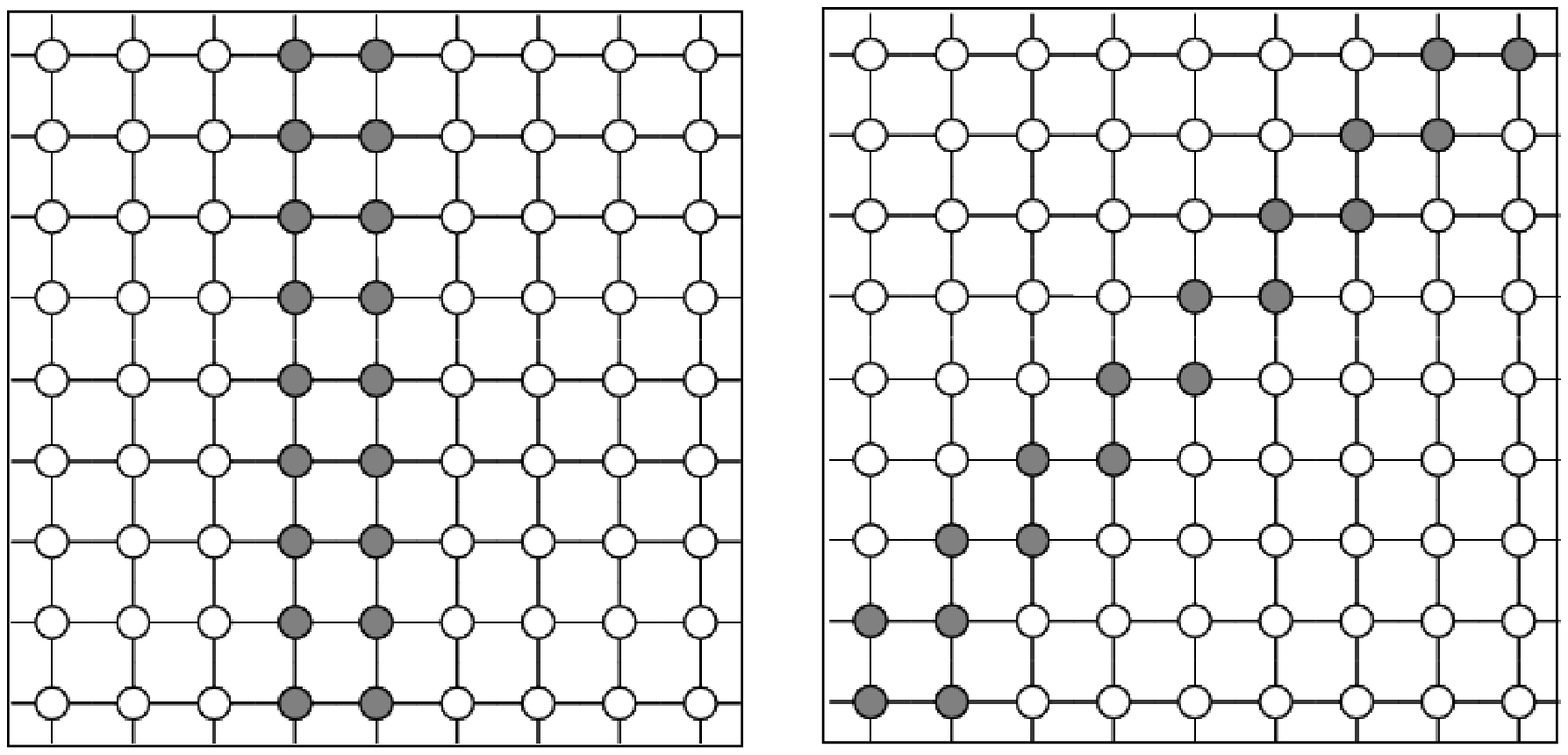}}\\
\subfigure[Spirallike wrapping clusters]{\includegraphics*[width=0.9\linewidth]{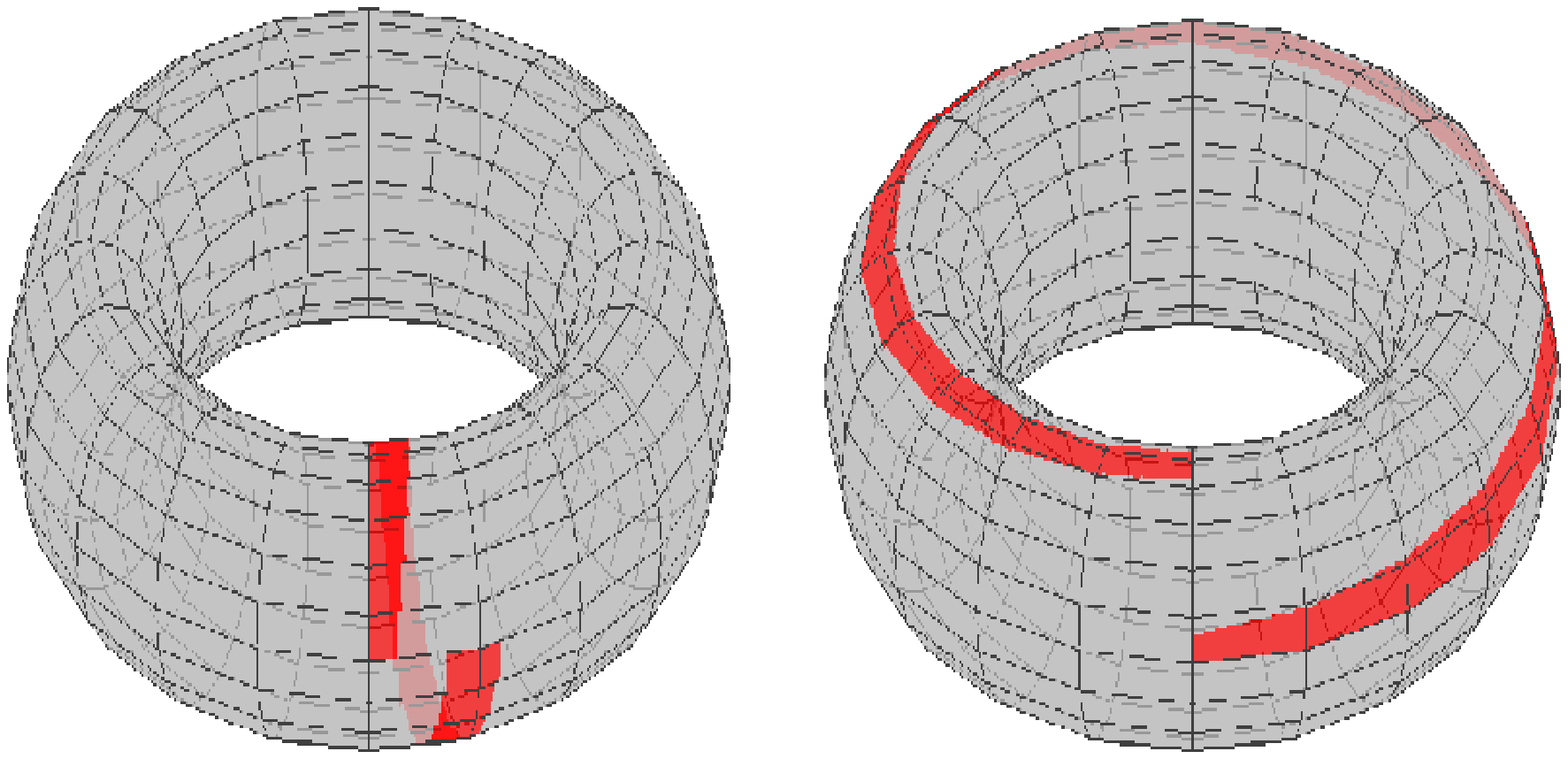}}\\
\subfigure[Ringlike wrapping clusters]{\includegraphics*[width=0.9\linewidth]{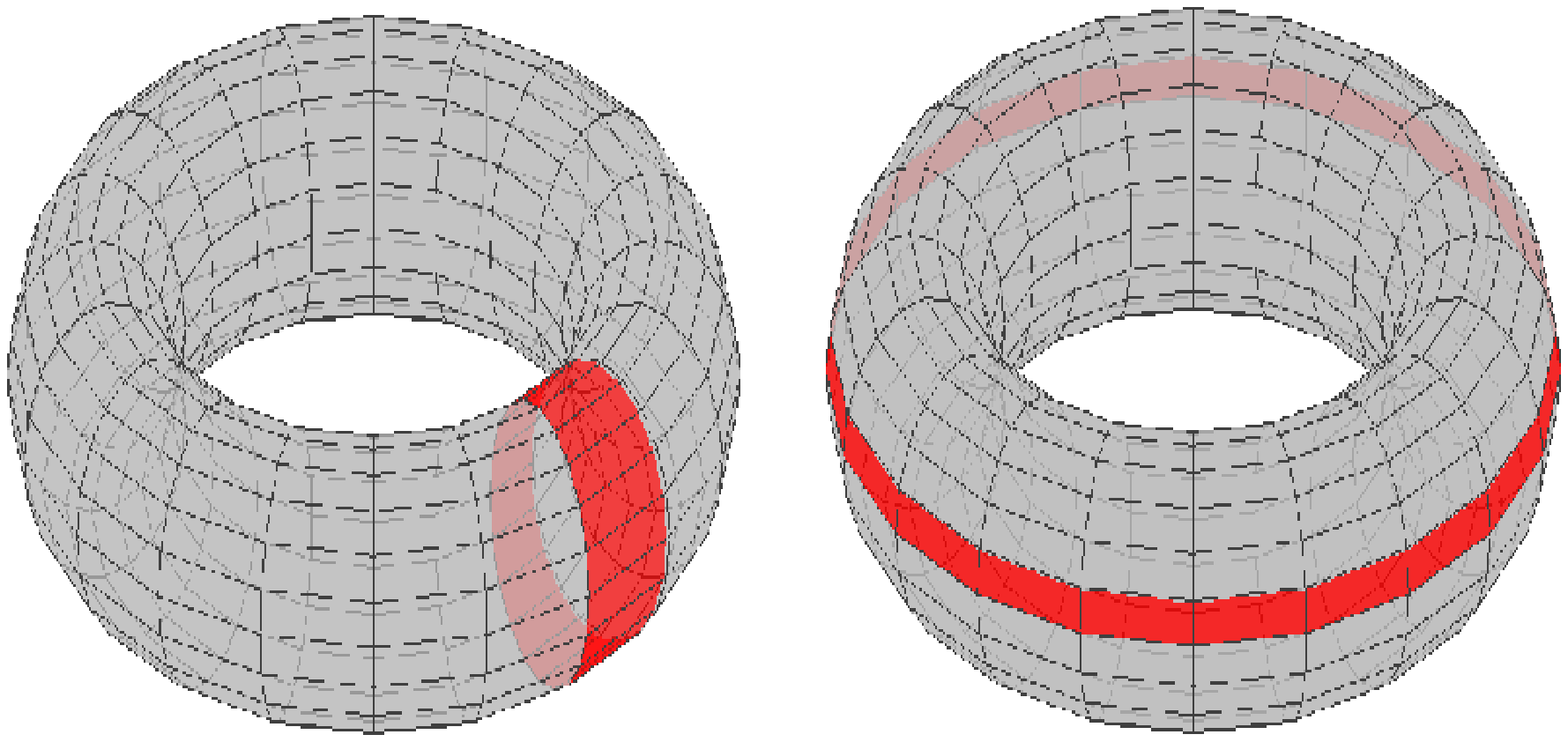}}\\
\caption {\label{fig:clusters}(Color online) Percolating clusters of different sorts on a plane and on a torus.}
\end{figure}

A crossing cluster is determined as a cluster that connects two opposite borders of lattice with open boundary conditions.  Examples of crossing clusters that percolate along vertical direction or simultaneously along vertical and horizontal directions are presented in Figure~\ref{fig:clusters}a.

A wrapping cluster is determined as a cluster which winds (i.e. provides a path of length $2\pi$) around the lattice with the periodic (toroidal) boundary conditions along the given direction~\cite{Pruessner2003}. The wrapping cluster may be either disconnected (spirallike) (Figure~\ref{fig:clusters}b) or  continuous (ringlike) (Figure~\ref{fig:clusters}c), or more complex one.

From the topological point of view, the spirallike clusters presented in Figure~\ref{fig:clusters}b are homotopic to a point, i.e, they can be continuously deformed to a point, and hence essentially differ from the ringlike clusters shown in Figure~\ref{fig:clusters}c. From the physical point of view, it is rather natural to think that applying periodic boundary conditions cannot destroy a percolating state existing in plane with open boundary conditions. Moreover, it can produce a new percolating state because of additional kind of symmetry, i.e. translation symmetry.

In our study, a system is considered as percolating if at least one spiral cluster (Figure~\ref{fig:clusters}b) can be found. For definiteness we call it as a problem of \emph{physical percolation on a torus} in contrast with \emph{topological percolation} when only self-connected clusters are treated as wrapping ones~\cite{Pinson1994JSPh}.

The value of threshold concentration may be determined by calculation of the probability  $R_L(p)$ for a cluster to cross a square lattice of $L \times L$ sites, if the boundary conditions are open, or to wrap around the periodic boundary conditions. In the thermodynamical limit  ($L \to \infty$), this probability is equal to the probability that the system percolates (i.e. it tends to the step-function and equals 0 below the percolation threshold and 1 above it)~\cite{Newman2000PRL}.

Since cluster wrapping can be defined in a number of different ways (see, e.g., \cite{Newman2000PRL}) there are a corresponding number of different probabilities $R_L$:
 \begin{enumerate}
 \item $R^{h}_L$ is the probability of wrapping horizontally around the system;
 \item $R^{v}_L$ is the probability of wrapping vertically around the system;
 \item $R_L^\text{or}$ is the probability of wrapping around either the horizontal or vertical direction, or both;
 \item $R_L^\text{and}$ is the probability of wrapping around both directions simultaneously.
\end{enumerate}

For the square lattices and isotropic problem these probabilities satisfy the following relations~\cite{Newman2000PRL,Newman2001PRE}:
\begin{equation}
R_L^{h}=R_L^{v},                               \label{eq:relationRhRv}\\
\end{equation}
\begin{equation}
R_L^{h}=   (R_L^\text{or} + R_L^\text{and})/2,             \label{eq:relationRh}\\
\end{equation}
as well as the inequalities
\begin{equation}
R_L^\text{and} \leq R_L^{h} \leq R_L^\text{or}. \label{eq:ineqRb}
\end{equation}
Relations \eqref{eq:relationRhRv}, and \eqref{eq:relationRh} evidence that only two of percolation probabilities  are independent. Obviously, for an anisotropic system the relation~\eqref{eq:relationRhRv} cannot hold and, hence, there are three independent probabilities. Nevertheless, for a strong anisotropic systems a spanning or wrapping cluster always arises along one direction, say vertical, and hence $R^v = R^\text{or}$, $R^h = R^\text{and}$.

The detailed studies have shown~\cite{Ziff1990JPhysA, Cornette2003} that for the specified problem (e.g., for crossing or wrapping clusters) and the criterion used, the curves $R_L(p)$ cross each other in a unique intersection point $R^*$ located at $p=p_c$ in the thermodynamical limit  ($L \to \infty$).

\begin{figure} [htbp]
\centering
\includegraphics[width=0.9 \linewidth]{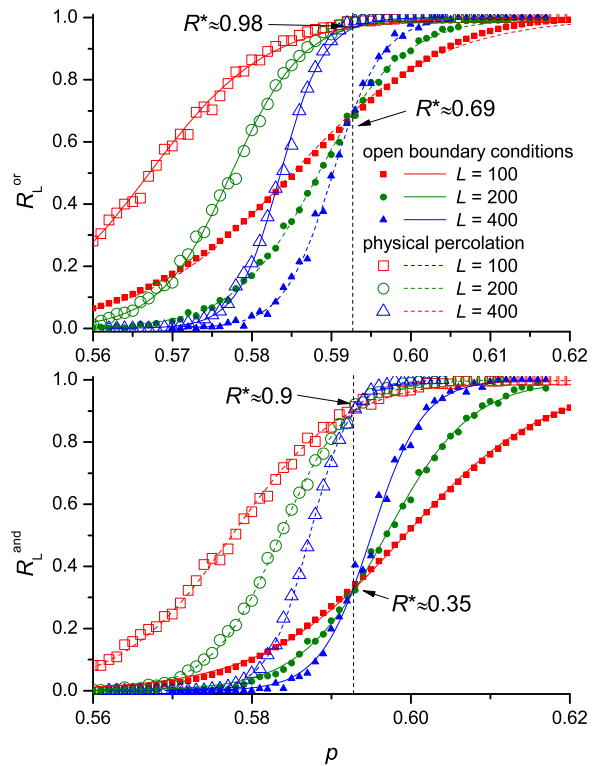}
\caption {\label{fig:comparison}(Color online)
Comparison of $R_L^\text{or}$ and $R_L^\text{and}$ versus $p$ dependencies for monomer problem ($k=1$) (physical percolation) and crossing clusters (with open boundary conditions) and different size of square lattice, $L$.
}
\end{figure}

Figure~\ref{fig:comparison} compares $R_L^\text{or}(p)$ and $R_L^\text{and}(p)$ dependencies for
monomer problem ($k=1$), i.e. conventional site problem and different size of square lattice, $L$.
The results are presented for the systems with periodic and open boundary conditions.

If the disconnected spiral clusters similar to shown in Figure~\ref{fig:clusters}b are not treated as percolating, the exact expressions of  $R^*$  at percolation threshold, $p_c$, for each of the definitions have been deduced~\cite{Ziff1990JPhysA} from the work by Pinson~\cite{Pinson1994JSPh}.  The values of $R^*$  presented by Newman and Ziff~\cite{Newman2000PRL,Newman2001PRE} are $R^{*\text{or}}= 0.690\,473\,725$, $R^{*\text{and}} = 0.351\,642\,855$.

The intersection points $R^*$ for physical percolation in our study are $R_L^{*\text{or}} \simeq 0.90$ and $R_L^{*\text{and}} \simeq 0.98$.

The $R_L(p)$ functions have been estimated by performing 1000 independent runs. Percolation concentration $p_c(L)$ for the lattice of given linear size $L$ filled with $k$-mers at the given concentration $p$ has been determined using the fitting function~\cite{Rintoul1997}
\begin{equation}
R_L(p) =\left(1 + \exp \left( -(p - p_c(L))a\right)\right)^{-1},
\label{eq:fit}
\end{equation}
where  $a$ is adjusted constant.

To extrapolate the estimations of the percolation thresholds $p_c(L)$ obtained at the lattice of size $L$ to the infinite large lattice $p_c(\infty )$, the usual finite size scaling analysis of the percolation behavior has been done. To perform extrapolation, we used at least three lattices of different sizes and scaling relation
\begin{equation}\label{eq:e2p}
    \left| p_c(L) - p_c(\infty ) \right|\propto L^{ - 1/\nu },
\end{equation}
where $\nu = 4/3$ is the critical exponent of correlation length for the 2d random percolation problem~\cite{Stauffer}. The universality of the $k$-mers problem has been justified earlier~\cite{Cornette2003}. In our study, the typical values of lattice size are $L=50k, 75k, 100k, 150k, 200k, 400k$.

\begin{figure*} [!htbp]
\centering
\subfigure[Isotropic case, $k = 2$, $s = 0.0$ ]{\includegraphics[width=0.33\linewidth]{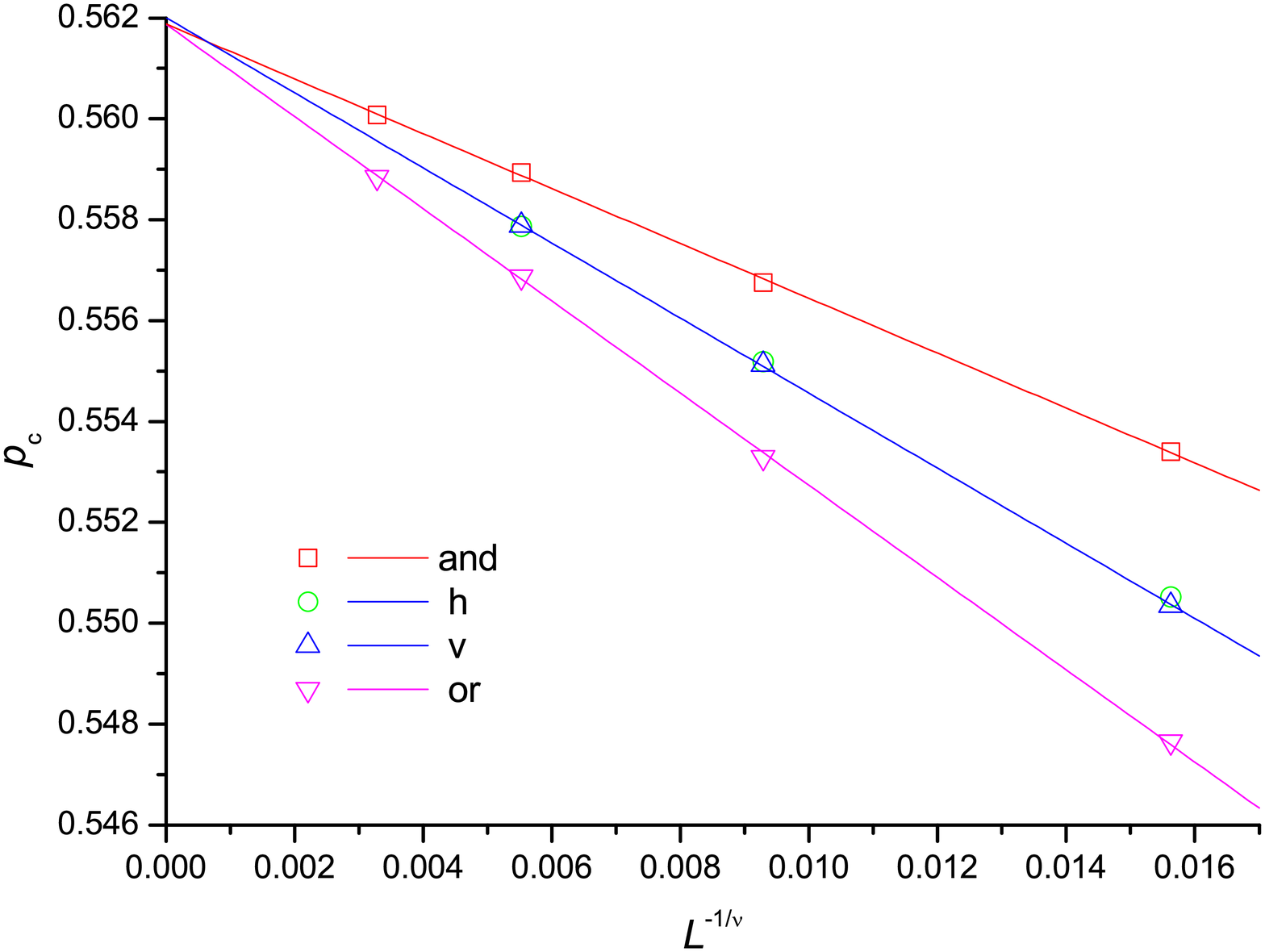}}\hfill
\subfigure[Slight anisotropic case, $k = 4$, $s = 0.1$ ]{\includegraphics[width=0.33\linewidth]{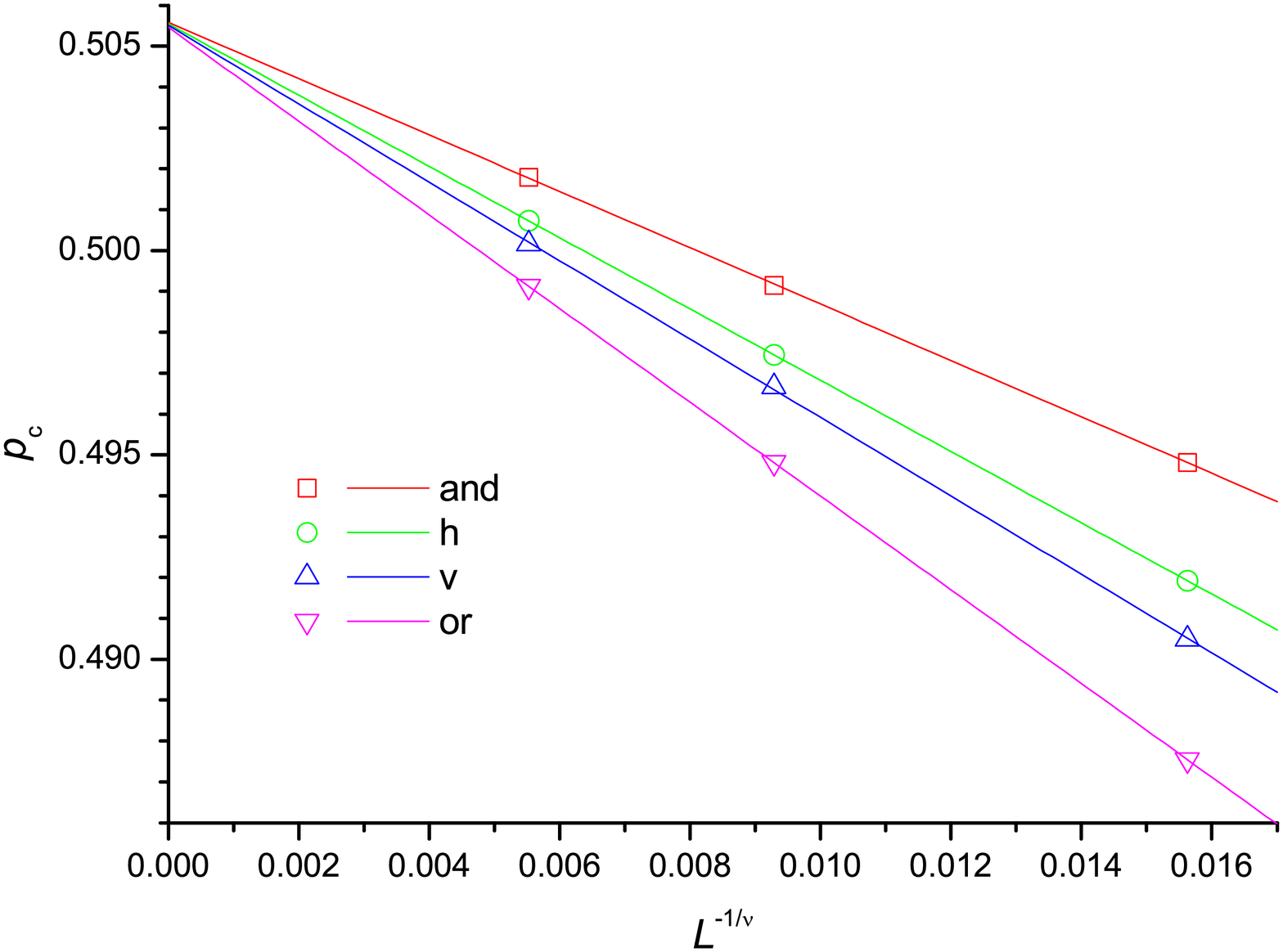}}\hfill
\subfigure[Anisotropic case, $k = 16$, $s = 0.8$ ]{\includegraphics[width=0.33\linewidth]{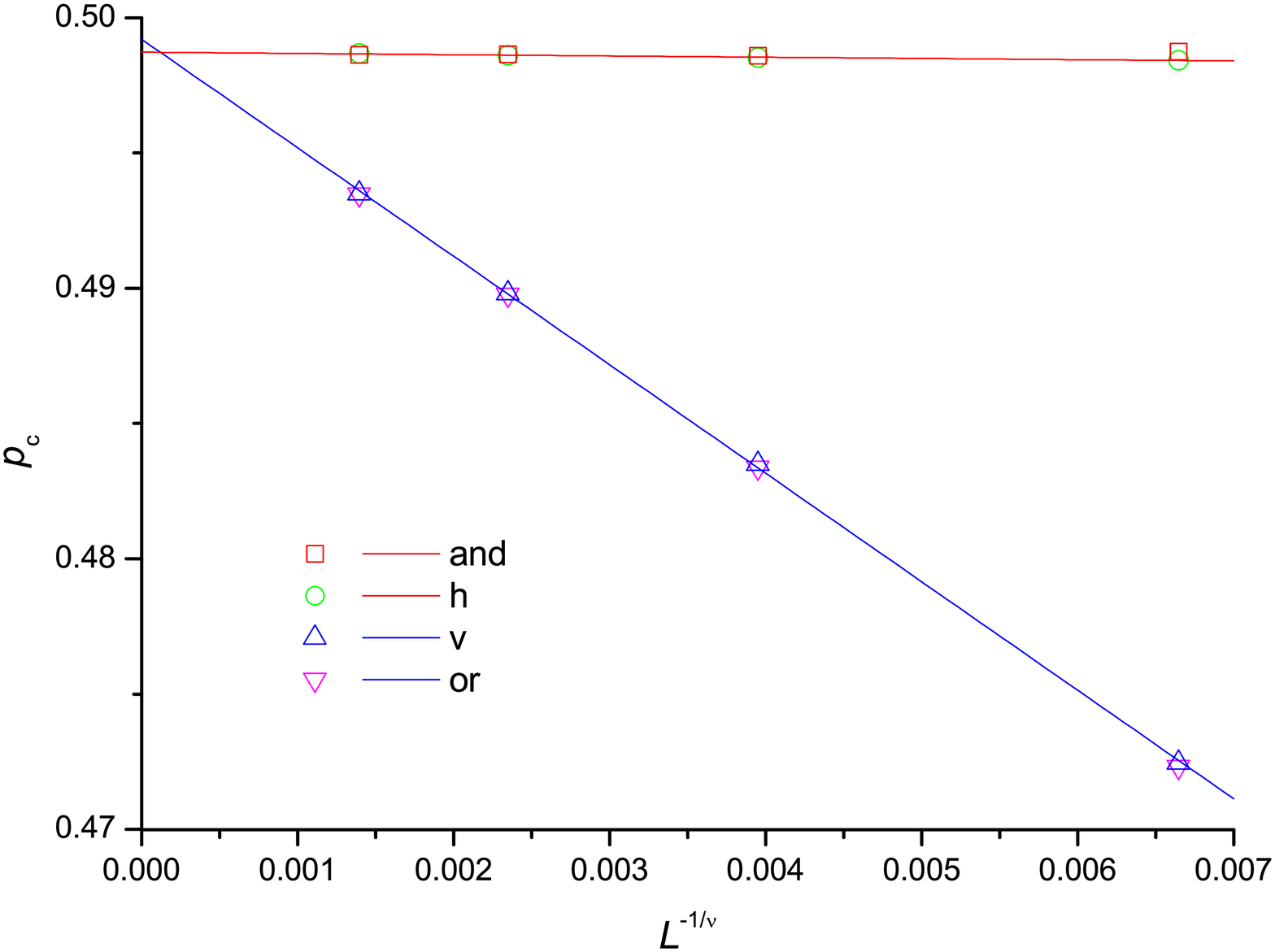}}
\caption {\label{fig:scaling16}(Color online) Percolation concentration $p_c$ versus size of the lattice $L$ for and different criteria for physical percolation.
}
\end{figure*}

Examples of $p_c$ versus $L$ scaling behavior for $k = 16$, $s = 0.8$ and
four criteria ($h,v, or, and$) are presented in Figure~\ref{fig:scaling16}.

The preliminary studies have shown that in all cases the $p_c(L)$  scaling is minimal for criterion $and$ and the final results on percolation concentration have been obtained using the criterium $and$.  To simplify the notation, below we omit superscript $and$  where it is possible.

To avoid very time-consuming computations with the lattices of huge size for $k=256$ and $s=0$ , we used only two relatively small lattices $L = 50k$ and $L = 75k$ and two different criteria, namely \emph{and} and \emph{or}. Intersection points (i.e. $p_c(\infty )$) extracted from \eqref{eq:e2p} for two different criteria are almost the same within error bar about 0.001.

Another special case is $L=512$, $s=0$. Only one lattice size $L=37k$ has been used for rough estimation of the percolation threshold. Percolation concentration has been calculated from the equation $R_L^\text{and}(p_c)=0.9$.

\subsection{\label{sec:model3}Other details}

Breadth-first search (BFS) algorithm has been applied to identify a percolation cluster. BFS seems to be faster and more appropriate for the toroidal boundary conditions than Hoshen--Kopelman (HK76) algorithm~\cite{Hoshen1976}. The additional tests have shown that results obtained using BFS and HK76 algorithms are identical within error bar.

The mean degree of the system anisotropy has been calculated as
\begin{equation}
\delta=\sum_{i=1}^{N_c}N_i\alpha_i/N_t
\label{eq:delta}
\end{equation}
where  $\delta_i=(R_i^y-R_i^x)/R_i$. Here, $R_i^y$ and $R_i^x$ are radii of gyration of cluster $i$ in $y$ and $x$ directions, respectively, $R_i$ is its mean radius of gyration, $N_c$ is a total number of clusters, $N_i$ is a number of filled sites in the cluster $i$, and $N_t$ is a total number of the filled sites.

\section{\label{sec:results}Results and discussion}

\subsection{\label{subsec:FinSizeScal} Non-universality of intersection points $R^*$}

\begin{figure}[!htbp] 
\centering
\includegraphics[width=\linewidth]{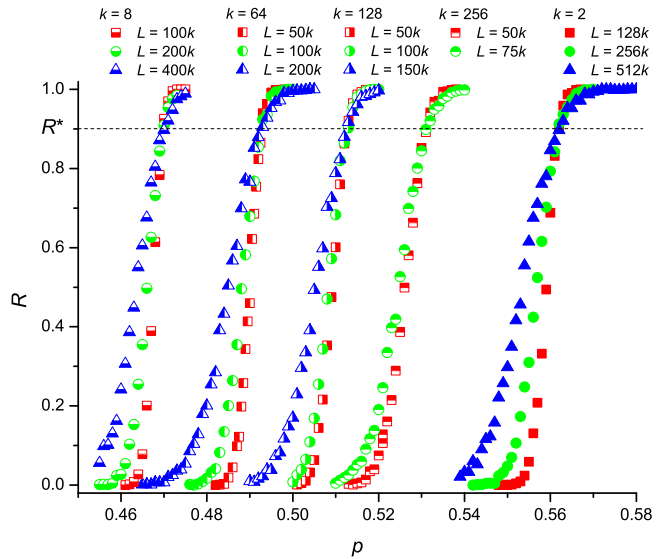}
\caption {\label{fig:RvsPS0}(Color online) Probability curves for isotropic systems, $s=0$,  and
different values of $k$ and $L$. Arrows indicate the intersection points.}
\end{figure}

The value of the percolation probability or percolation cumulant at the  intersection point $R^*$ may be important characteristic representing the universality class~\cite{Longone2012}. Figure~\ref{fig:RvsPS0} presents examples percolation probability $R_L$ versus $k$-mers concentration $p$
for isotropic systems, $s=0$, and different values of $k$ and $L$.

For isotropic problem the position of intersection point remained unchanged within precision of estimation, being $R^*\simeq 0.90$ for all $k$ within the interval between 1 and 512. This behavior is rather similar to that observed for percolation problem of $k$-mers with open boundary condition~\cite{Cornette2003}. For the criterium $and$, the same values of $R^{*}\simeq 0.3$ have been observed for the different length of $k$-mers ranging between $k = 1$ and  $k = 25$. Thus, universality of intersection points $R^*$ has been observed for the systems with different boundary conditions (periodical and open) and it may indicate the conserving of universality class irrespective of the size of $k$-mers.

\begin{figure}[!htbp] 
\centering
\includegraphics[width=\linewidth]{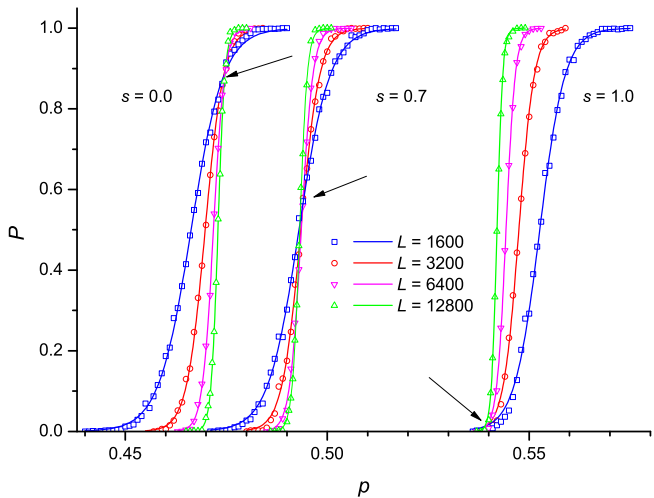}
\caption {\label{fig:RvsP}(Color online) Probability curves for $k=32$, $s=0.0, 0.7, 1.0$. Arrows indicate the intersection points.}
\end{figure}

However, such universality of intersection points $R^*$ has been not observed for anisotropic systems. Figure~\ref{fig:RvsP} presents examples of percolation probability $R$ versus $k$-mer concentration $p$
for $k=32$ and different values of $s$ and $L$. At fixed value of $k$ the position of intersection point $R^*$ continuously decreased  with increasing of $s$.

\begin{figure}[!htbp] 
\centering
\includegraphics[width=\linewidth]{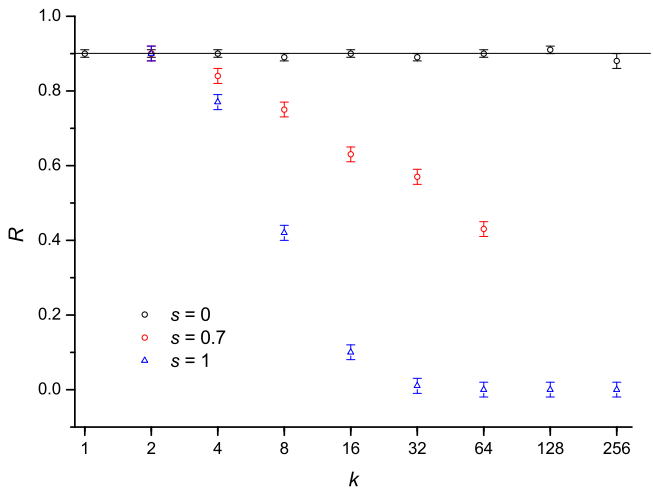}
\caption {\label{fig:Rvsk}(Color online) Intersection point of percolation probability $R^*$ versus $k$ at $s=0.0, 0.7, 1.0$.}
\end{figure}

The more detailed studies have shown that for anisotropic systems the position of intersection points $R^*$ also  dependents  on the value of $k$ (Figure~\ref{fig:Rvsk}). For the completely ordered systems, $s=1$, the value of $R^{*}$ decreased monotonically and became close to 0 for larger sizes of $k$-mers.
This observation may reflect the continuous change of universality class and correspondences to previously reported data for the completely ordered systems with open boundary conditions~\cite{Longone2012}. For partially ordered systems, the similar effect of $k$-mers length on the value of $R^{*}$ has been observed (Figure~\ref{fig:Rvsk}).

\begin{figure*}[!htbp]
\centering
\subfigure[$k=2$]{\includegraphics*[width=0.32\linewidth]{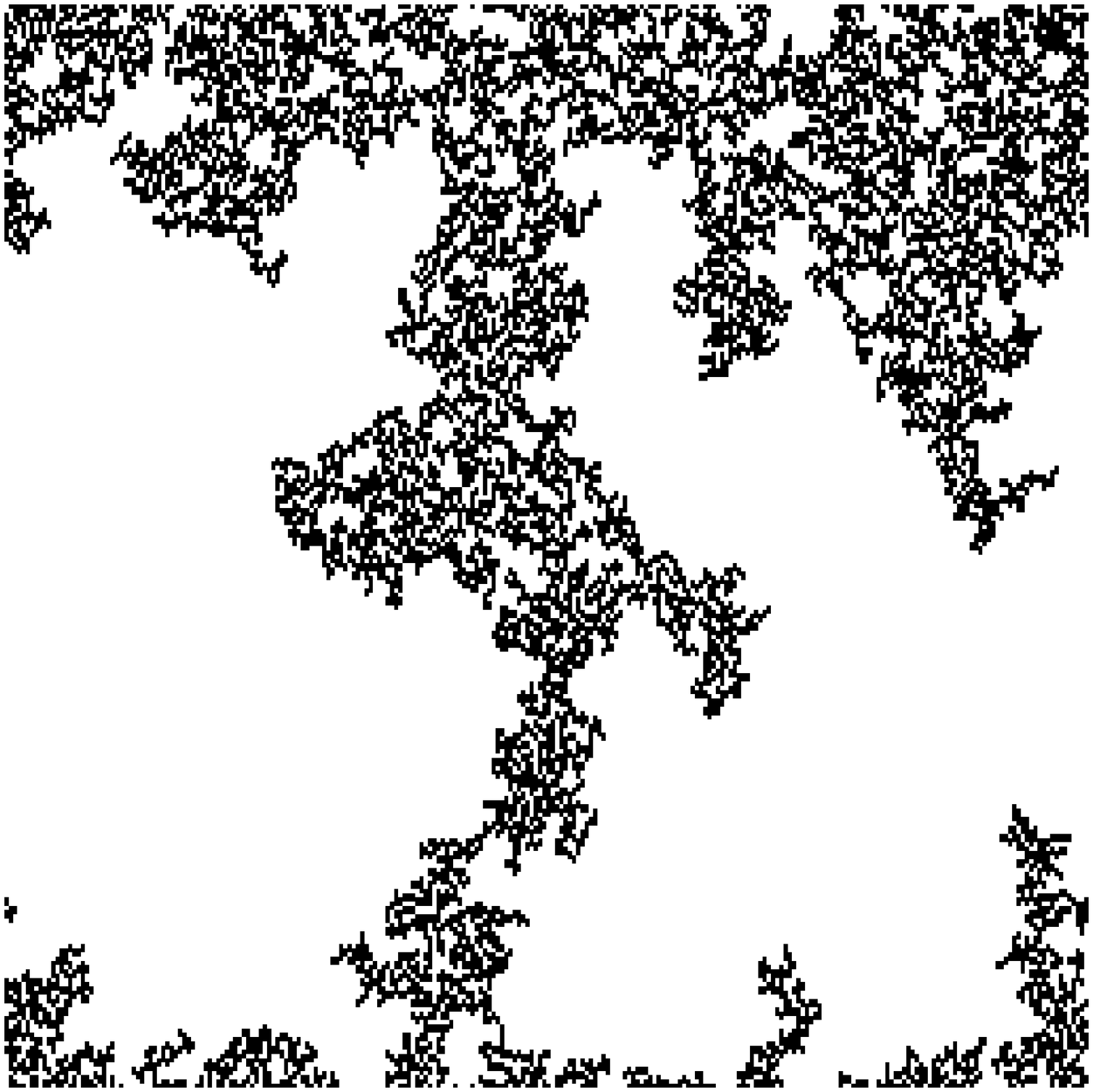}}\hfill
\subfigure[$k=8$]{\includegraphics*[width=0.32\linewidth]{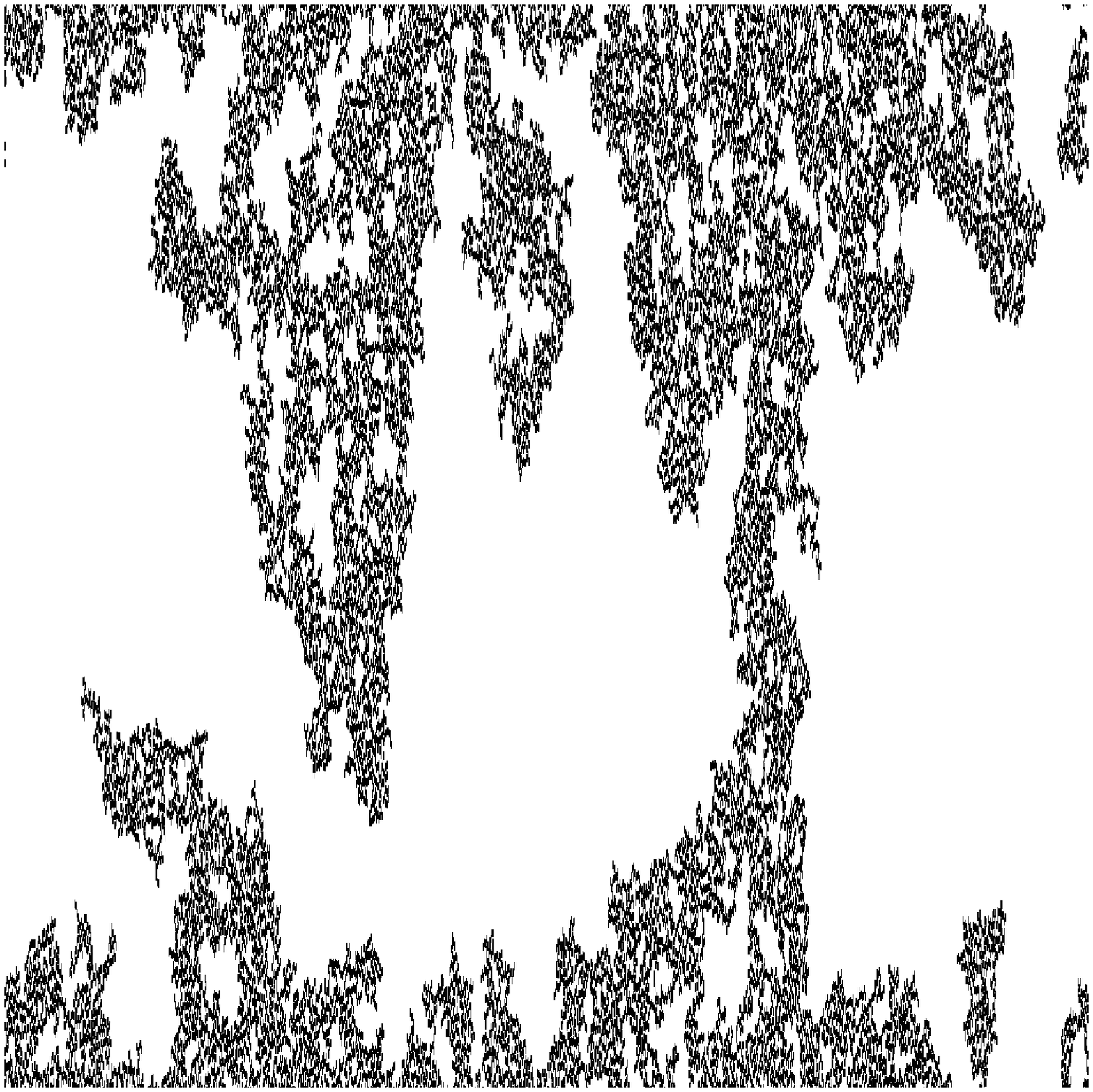}}\hfill
\subfigure[$k=32$]{\includegraphics*[width=0.32\linewidth]{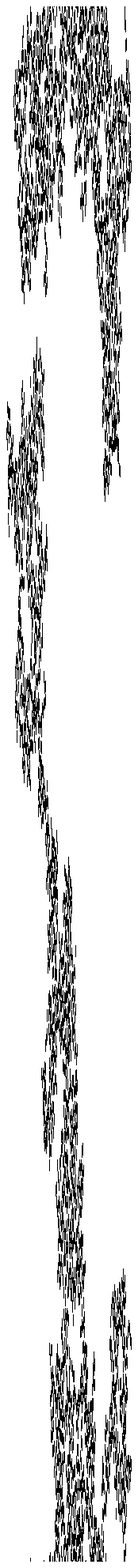}}\\
\caption{\label{fig:patternS10} Examples of wrapping clusters incipient in vertical direction  for completely ordered system ($s=1.0$) for different length of $k$-mers. The size of a square lattice is $L=128k$. Periodical boundary conditions.}
\end{figure*}

Thus, orientation of $k$-mers affected the universality class of this percolation problem and it has been conserved only for the isotropic systems ($s=0$), where universality is the same for the different length of $k$-mers. It can be speculated that this violation of universality can reflect the effect of the system anisotropy. This anisotropy has been maximally denominated for the completely ordered systems ($s=1$) where the effect of the $k$-mer length on the  value of $R^*$ is maximal (Figure~\ref{fig:Rvsk}). The more detailed analysis shown that the structure of percolation clusters is strongly depends upon $k$, they have been elongated along vertical direction and degree  of elongation increased as length of $k$-mers increased (Figure~\ref{fig:patternS10}). Moreover, the mean degree of  system anisotropy $\delta$ calculated using Eq.~\ref{eq:delta} is dependent on $k$-mer concentration $p$, length $k$ and order parameter~$s$.

\begin{figure*}[!htbp]
\centering
\subfigure[]{\includegraphics*[width=0.45\linewidth]{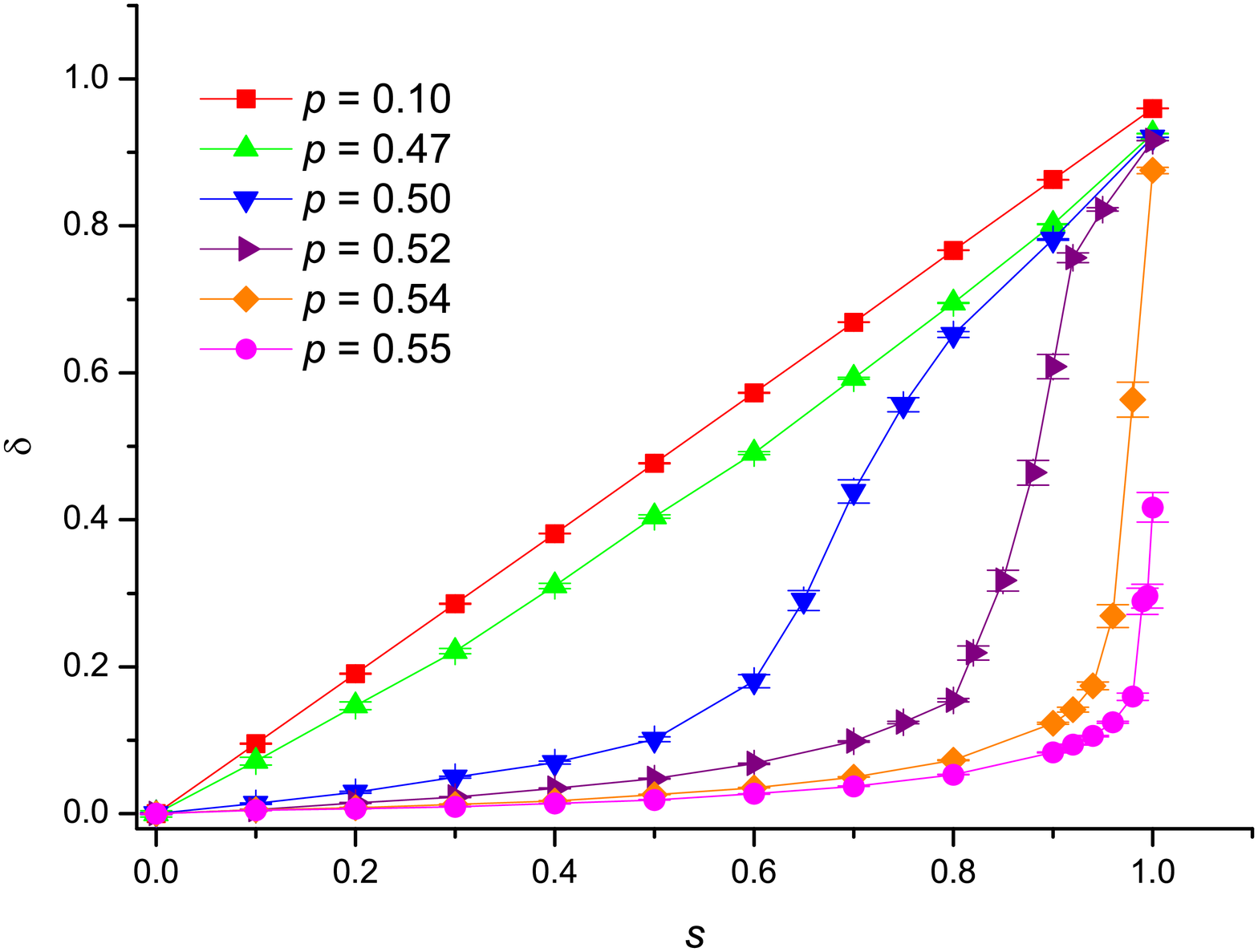}}\hfill
\subfigure[]{\includegraphics*[width=0.45\linewidth]{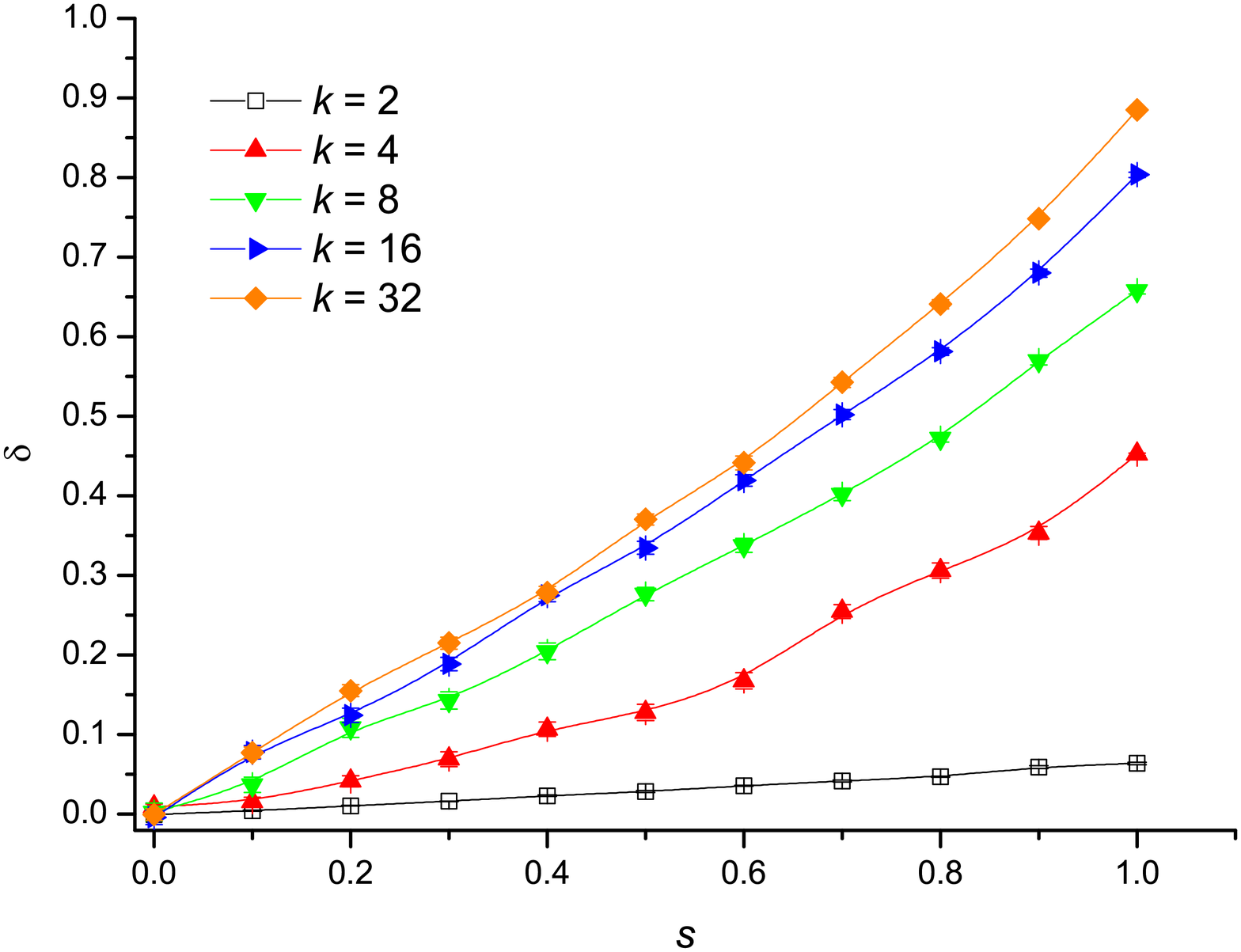}}\hfill
\caption{\label{fig:delta}
Mean degree of the system anisotropy $\delta$ versus order parameter $s$: a)
at different concentrations $p$ and the fixed length of a $k$-mer,  $k=32$; b)
at different $k$-mer length $k$ and concentrations that have been corresponded to the percolation transition for given systems. The size of lattice is $L=4096$ and the data have been averaged on 100 independent runs.
}
\end{figure*}

Figure~\ref{fig:delta}a presents examples of $\delta$ versus order parameter $s$ at different fixed concentrations $p$ and fixed length of $k$-mer, $k=32$. The size of lattice is relatively large, $L=4096$, so, the finite size effects are rather small. For isotropic systems (at $s=0$), the value of $\delta$ is always zero and it is maximal for completely ordered systems (at $s=1$). At small values of $p$ the relation between $\delta$ and $s$ is nearly linear. With increasing of $p$ and fixed $s$ the  value of $\delta$ decreased, however, it noticeably dropped above percolation threshold and became practically zero in the vicinity of jamming concentration. E.g., the concentration of $p=0.50$ is above the percolation threshold for the systems with order parameter $s$ below $\simeq 0.8$ and here, the $\delta(s)$ dependence noticeably deviates from near linear (Figure~\ref{fig:delta}a). Figure~\ref{fig:delta}b presents examples of $\delta$ versus order parameter $s$ at different fixed length of $k$-mer, $k=32$ and the concentrations that has corresponded to the percolation transitions for the given systems. The value of $k$-mer length $k$ strongly affected the mean degree of the system anisotropy $\delta$ at the percolation transition. E.g.,  for dimers, $k=2$, the value of  $\delta$ is rather small in the whole range of $s$ between $0$ and $1$, however, with increasing of $k$, the $\delta (s)$ became more noticeable and we believe that they can transfer into the near-linear of type  $\delta\simeq s$ in the limit of large $k$-mer length, $k\to\infty$.

\subsection{\label{subsec:ptin} Dependence of percolation threshold $p_c$ versus order parameter $s$}

\begin{figure}[!htbp] 
\centering
\includegraphics[width=\linewidth]{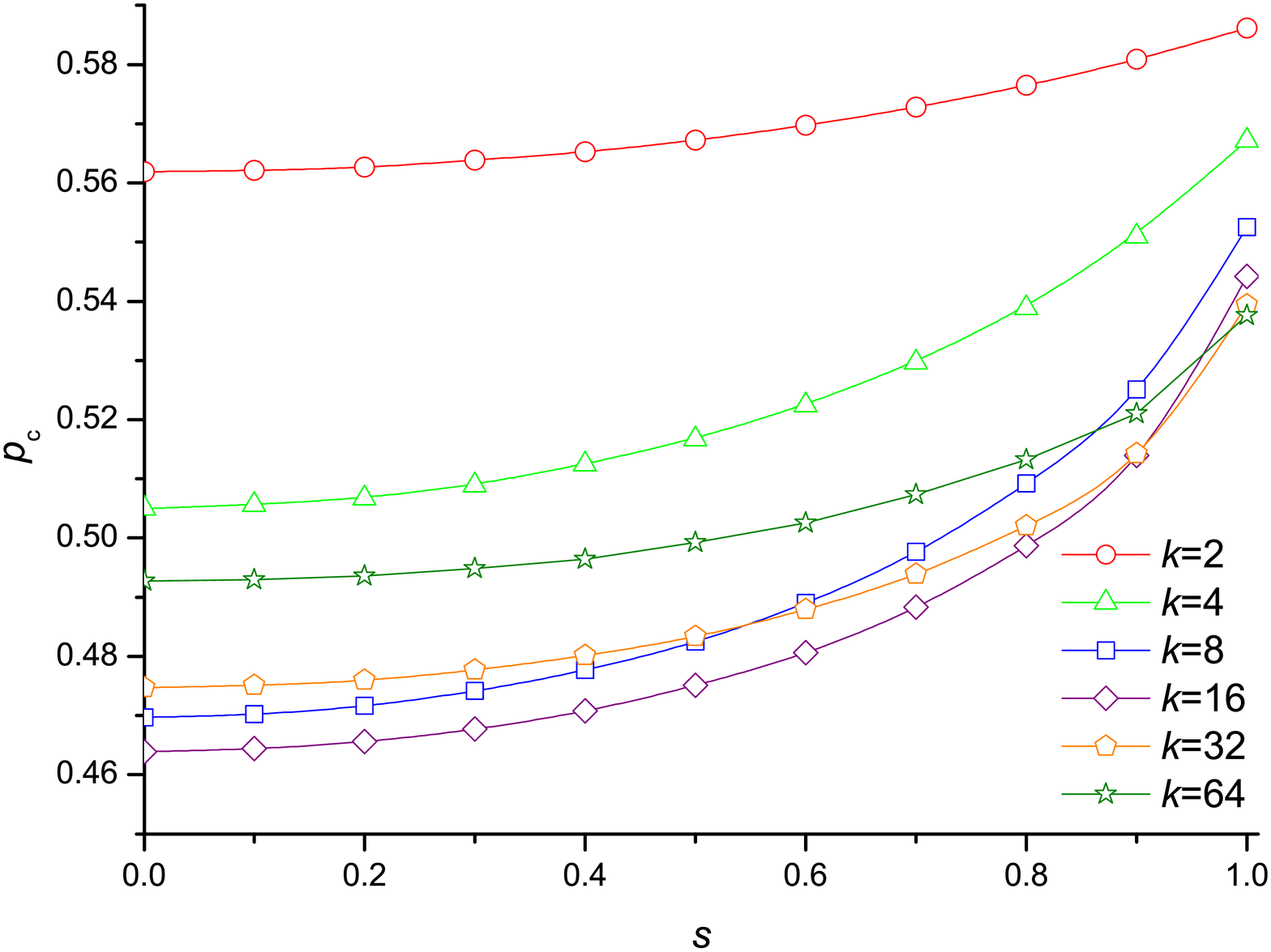}
\caption {\label{fig:pc_s}(Color online) Percolation threshold $p_c$ versus order parameter $s$ for $k$-mers of different length.
}
\end{figure}

The $p_c(s)$ dependencies for $k$-mers of different length ($k=2 \dots 128$ are presented in Figure~\ref{fig:pc_s}. For completeness, the precise numerical information is also collected in Table~\ref{tab:pc}. In addition the Table~\ref{tab:pcestimation} presents rougher estimations for $k=256$ and $k=512$ for isotropic ($s=0$) and completely ordered ($s=1$) systems.

\begin{table*}[!htbp]
  \caption{Percolation threshold $p_c$ versus order parameter $s$ for $k$-mers of different length $k$.}\label{tab:pc}
\begin{ruledtabular}
  \begin{tabular}{cccccccc}
$s$ &  $k=2$  &  $k=4$  &  $k=8$  &  $k=16$ &  $k=32$ &  $k=64$ &  $k=128$ \\
\hline
0.0 &  0.5619 &  0.5050 &  0.4697 &  0.4638 &  0.4748 & 0.4928 & 0.5115\\
0.1 &  0.5621 &  0.5056 &  0.4702 &  0.4644 &  0.4751 & 0.4930 &    \\
0.2 &  0.5627 &  0.5067 &  0.4717 &  0.4656 &  0.4763 & 0.4936 &   \\
0.3 &  0.5638 &  0.5090 &  0.4742 &  0.4677 &  0.4777 & 0.4948 &   \\
0.4 &  0.5653 &  0.5124 &  0.4777 &  0.4708 &  0.4802 & 0.4964 &   \\
0.5 &  0.5672 &  0.5167 &  0.4825 &  0.4751 &  0.4834 & 0.4993 &   \\
0.6 &  0.5698 &  0.5224 &  0.4890 &  0.4807 &  0.4879 & 0.5025 &   \\
0.7 &  0.5728 &  0.5296 &  0.4977 &  0.4883 &  0.4939 & 0.5074 &   \\
0.8 &  0.5765 &  0.5389 &  0.5092 &  0.4987 &  0.5021 & 0.5132 &   \\
0.9 &  0.5809 &  0.5510 &  0.5251 &  0.5140 &  0.5142 & 0.5210 &   \\
1.0 &  0.5862 &  0.5672 &  0.5526 &  0.5442 &  0.5397 & 0.5376 & 0.5366  \\
  \end{tabular}
  \end{ruledtabular}
\end{table*}

\begin{table}[!htbp]
  \caption{Estimations of percolation threshold $p_c$ for $k$-mers of ladge length $k$.}\label{tab:pcestimation}
\begin{ruledtabular}
  \begin{tabular}{ccc}
 $s$ & $k=256$&  $k=512$ \\
\hline
0.0 & 0.530 & 0.5485 \\
1.0 & 0.535 &  \\
  \end{tabular}
  \end{ruledtabular}
\end{table}

The obtained data evidence that the increase of system ordering always results in increase of $p_c$ value. Such behavior correlates with theoretical data obtained in the systems of partially oriented penetrable rods~\cite{Balberg1983,Balberg1983a,Boudville1989, Natsuki2005} and experimentally studied effect of carbon nanotube alignment on percolation in polymer composites~\cite{Du2005}.

\begin{figure} [!htbp]
\centering
\includegraphics[width=\linewidth]{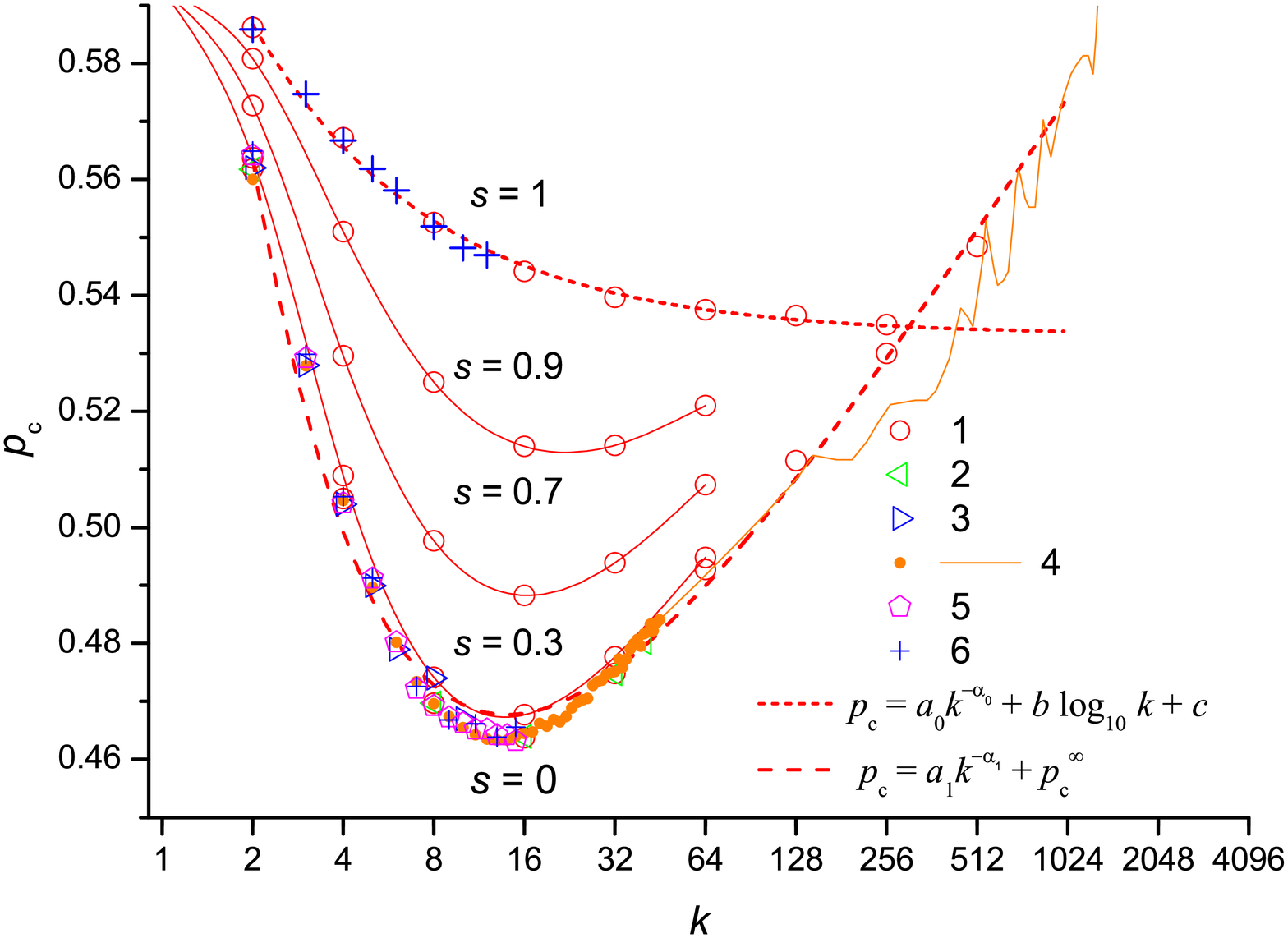}
\caption {\label{fig:pc_k}(Color online) Percolation threshold $p_c$ versus $k$-mer length $k$ at different values of order parameter $s$. Here, the different data are presented that have been obtained in: (1) this work, (2)~\cite{Leroyer1994}, (3)~\cite {Vandewalle2000}, (4)~\cite{Kondrat2002}, (5)~\cite{Cornette2003} and (6)\cite{Longone2012}. The dashed lines have been obtained by least square fitting of the data points using the Eqs.~\ref{eq:fit1},\ref{eq:fit2}.
}
\end{figure}

Figure~\ref{fig:pc_k} presents examples of $p_c$ versus $k$ dependencies obtained for different values of $s$ in this work (1), as well as data presented earlier for the isotropic ($s=0$) and completely ordered ($s=1$): (2)~\cite{Leroyer1994}, (3)~\cite {Vandewalle2000}, (4)~\cite{Kondrat2002}, (5)~\cite{Cornette2003} and (6)\cite{Longone2012}.

For completely ordered systems, i.e., at $s=1$, the percolation threshold $p_c$ monotonically decreased as value of $k$ increased. Recently, the similar behavior for $k$ from 1 to 12 with the asymptotic limit of $p_c^\infty=p_c(k \to \infty)\simeq 0.54$ has been  reported~\cite{Longone2012}.

The analysis have shown that the data obtained in our work may be rather well fitted by the power function
\begin{equation}
p_c=a_1/k^{\alpha_1}+p_c^\infty,
\label{eq:fit1}
\end{equation}
where  $p_c^*=0.533 \pm 0.001$, $a_1=0.088 \pm 0.003$, $\alpha_1=0.72 \pm 0.04$ and $r^2=0.998$ for the coefficient of determination.

Note, that for completely ordered penetrating anisotropic objects and continuous problem the excluded volume theory predicts the absence of noticeable dependence of the percolation threshold on aspect ratio $k$~\cite{Balberg1983,Balberg1983a}. In our lattice problem, observed effect of $p_c(k)$ dependence may reflect the influence of the lattice discreteness on the percolation threshold.

In our problem at $s=1$, the formation of percolation cluster reflects the mode of connectivity between vertically oriented one dimensional chains of $k$-mers. It may be assumed that in the limit of $k\to\infty$ the connectivity of two $k$-mers in the neighbor vertical lines at their end sites is sufficient for a formation of percolation cluster with minimal concentration of $p=0.5$, that is close to the numerically estimated value of
$p_c^*=0.533 \pm 0.001$.

In contrast, for partially ordered systems, i.e., at $s<1$, the percolation threshold $p_c$ is a nonmonotonic function of $k$ and for a certain length of $k$-mers $k=k_m$ a minimum of $p_c$ has been observed (Figure~\ref{fig:pc_k}). In total, the data obtained in this work for isotropic systems (i.e., at $s=0$)  have been in good correspondence with previously published data~\cite{Leroyer1994,Vandewalle2000,Kondrat2002,Cornette2003,Longone2012} with only exception to those obtained for the very long $k$-mers ($k>64$) in~\cite{Kondrat2002}. This inequality may reflect the relatively moderate size of lattices  that has been used in~\cite{Kondrat2002} ($L\leq  2500$) whereas in our simulations the maximum size of the lattice $L \sim 100k$, as a rule. In any case, our data confirmed the conclusion by Leroyer and Pommiers~\cite{Leroyer1994} and Kondrat and P\c{e}kalski~\cite{KondratPre63} about the presence of minimum at the $p_c$ versus $k$ dependence. For the disordered systems the position of the minimum, $k_m$, is dependent on the value of $s$, e.g., it is $k_m\simeq 13$ at $s=0$, $k_m\simeq 16$ at $s=0.7$, $k_m\simeq 22$ at $s=0.9$, and it seems that $k_m \to \infty$  in the limit of $s \to 1$ (Figure~\ref{fig:pc_k}). Note, that the asymptotic limit of $p_c^*=p_c(k\to\infty)\simeq 0.461$ derived in~\cite{Cornette2003} for $s=0$ in fact is very close to the value of $p_c$ at point of minimum, $k_m\simeq 13$.

\begin{figure*}[!htbp]
\centering
\subfigure[]{\includegraphics*[width=0.32\linewidth]{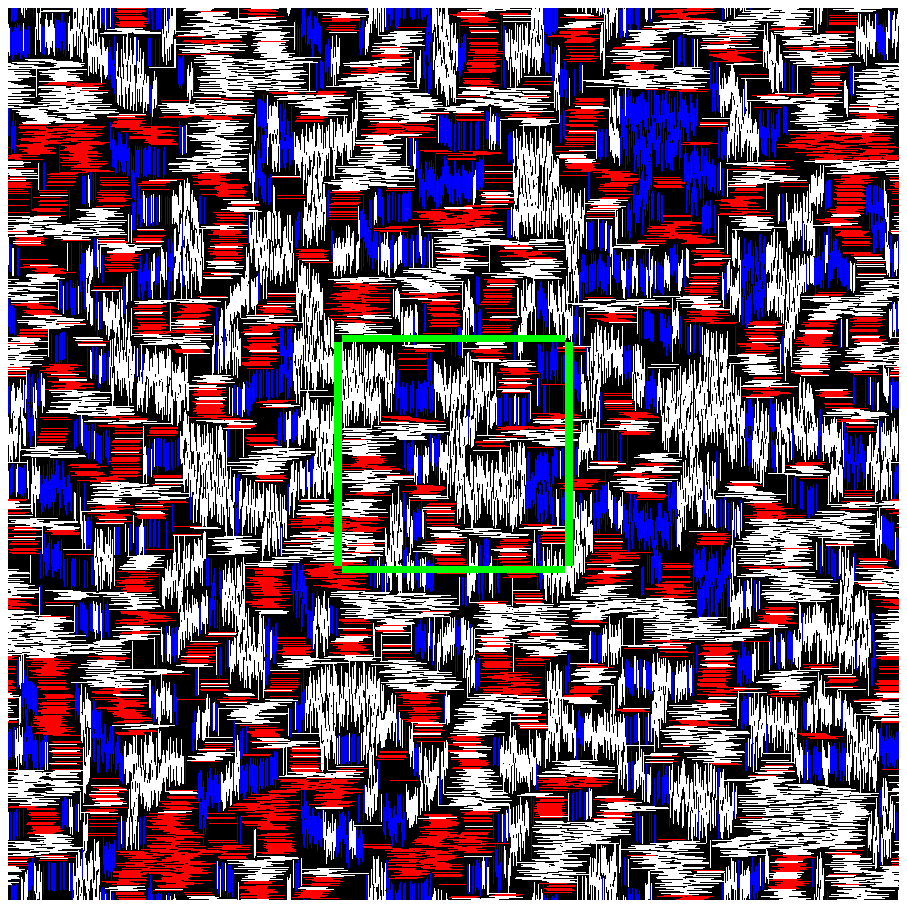}}\hfill
\subfigure[]{\includegraphics*[width=0.32\linewidth]{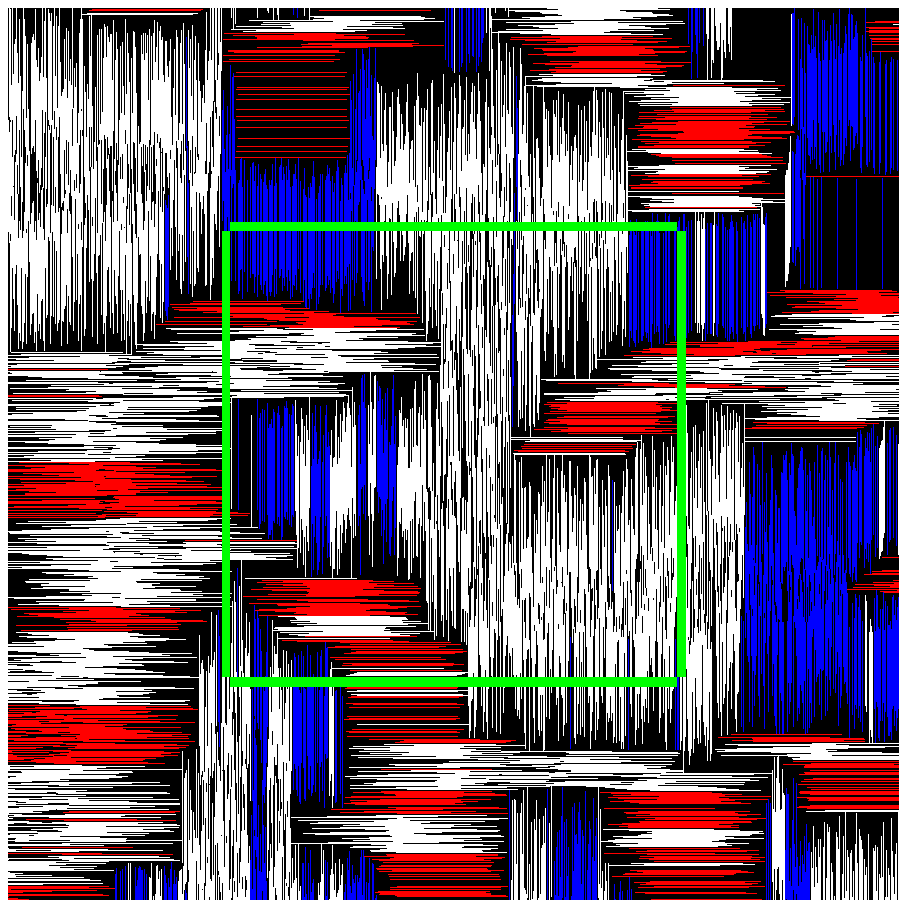}}\hfill
\subfigure[]{\includegraphics*[width=0.32\linewidth]{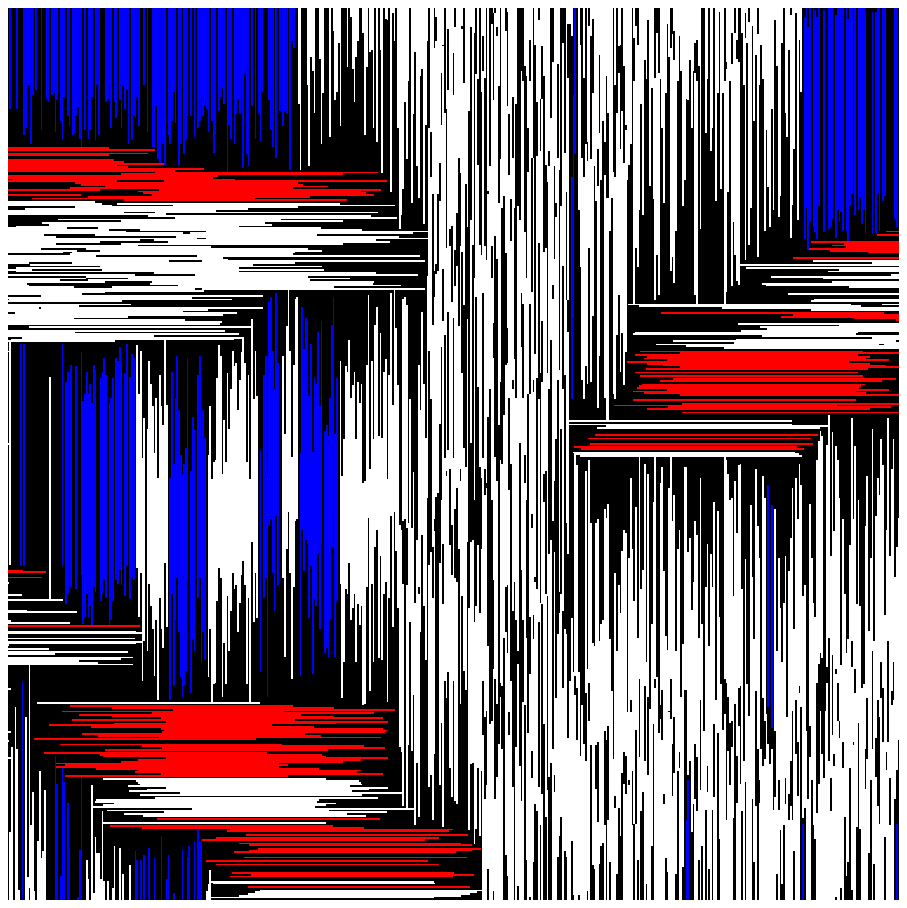}}\\
\caption{\label{fig:blocks} (Color online) Examples of percolation configurations of $k$-mers ($k = 128$) on a square lattice of size $L = 4096$ at $s=0$. Vertical and horizontal orientations are represented by different gray levels in printed version and in red and blue in online
version; empty sites are labeled black and sites of percolation cluster are labeled white. Here, (a) shows the whole lattice; (b)  shows the magnification of the pattern (a) in the central square; (c) shows the magnification of the pattern (b) in the central square.}
\end{figure*}

It is attractive to speculate that extremal $p_c$ versus behavior for partially ordered systems may reflect the competition of the two different effects influencing the value of percolation threshold.
We tried to fit the obtained data for isotropic system ($s=0$) using the function
\begin{equation}
p_c=a_0/k^{\alpha_0}+b\log_{10} k+ c
\label{eq:fit2}
\end{equation}
and had obtained the following numerical estimations for the parameters $a_0=0.36 \pm 0.02$, $\alpha_0=0.81\pm 0.12$, $b=0.08 \pm 0.01$, $c=0.33 \pm  0.02$,  and $r^2=0.991$ for the coefficient of determination.

It is remarkable, that exponents $\alpha_0=0.81\pm 0.12$ and $\alpha_1=0.72 \pm 0.04$ are practically the same for $s=0$ and $s=1$, respectively, and it may reflect the same effect of the discreteness on the percolation at the relatively small $k$ ($\leq10$). For disordered systems, the logarithmic increase of $p_c$ at large values of $k$ (Eq.\ref{eq:fit2}) may reflect the tendency of $k$-mers for stacking, or formation of squarelike blocks, especially  at large values of $k$. Such blocks of vertically and horizontally oriented $k$-mers are typical for partially ordered systems in jamming configurations~\cite{Lebovka2011PRE}, however, they are also important  at the percolation threshold. Examples of $k$-mer patterns ($k=128$) in the percolation point are presented in Figure \ref{fig:blocks} for isotropic system ($s=0$). The sequential magnification of the system has shown the  presence of rather compact blocks of vertically and horizontally oriented $k$-mers that have been connected into the percolating structure by overhanging of $k$-mers. The numerical studies have shown that for the ideal blocks, i.e. $k \times k$ squares the percolation concentration increased and jamming concentration decreased as $k$ value increased~\cite{Nakamura1987} and above certain critical value of $k$ no percolation has been observed. In this situation even at the saturation coverage (jamming) where no more object can be placed without any overlap there exist only finite clusters of $k \times k$ squares. We can assume the similar mechanism that governs the observed $p_c\propto k$ increasing of percolation threshold.

\begin{figure} [!tbhp]
\centering
\includegraphics[width=\linewidth]{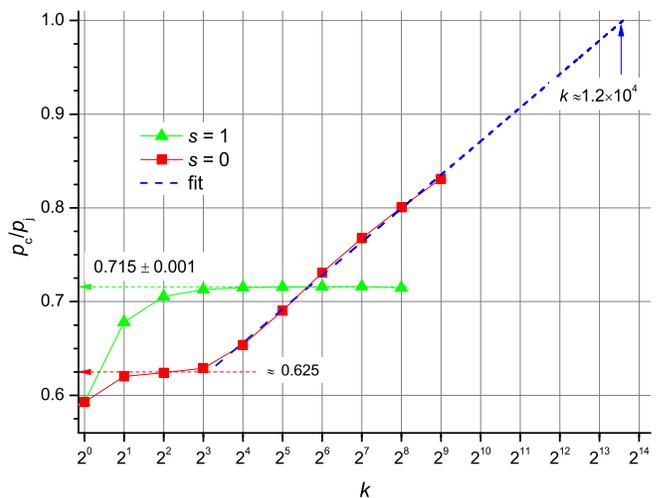}
\caption {\label{fig:pcpj}(Color online) Ratio of percolation and jamming concentration $p_c/p_j$ versus $k$-mer length $k$ for disordered ($s=0$) and completely ordered $s=1$ systems. The dashed lines for $s=0$ has been obtained by least square fitting of the data points using the Eqs. \ref{eq:fit3}.
}
\end{figure}

We checked the validity of conjecture of Vandewalle et al.~\cite{Vandewalle2000} about the constancy of the ratio of percolation and jamming concentration $p_c/p_j$ for disordered ($s=0$) and completely ordered $s=1$ systems (Figure~\ref{fig:pcpj}). The values of jamming concentration $p_j$ have been taken from our previously published work~\cite{Lebovka2011PRE}. For completely ordered systems ($s=1$) this ratio initially increased and became practically constant, $p_c/p_j\simeq 0.715$, at relatively large length of $k$-mers, $k>8$.

For isotropic systems ($s=0$) this ratio is approximately constant only small values of
$k$ ($k=2 \dots 8$) and for larger $k$, $k=16 \dots 256$  the $p_c/p_j$ increased proportionally to $\log_{10} k$:
\begin{equation}\label{eq:fit3}
p_c/p_j = b  \log_{10} k + c,
\end{equation}
where $b=0.119 \pm 0.003$ and $c = 0.513 \pm 0.006$ are the constants.

Thus, the constancy of the ratio $p_c/p_j$ fulfills for the completely ordered systems ($s=1$). It may reflect the similar influence of the discreteness of the lattice on the both  jamming and percolation.
On the other hand, the non-constancy of the ratio $p_c/p_j$ for isotropic systems ($s=0$) may reflect the different influence of the stacking on the jamming and percolation. For this case the approximation of the data presented in Figure \ref{fig:pcpj} gives $p_c/p_j\simeq 1$ at $k\simeq 1.2\times10^4$. So, we can suppose that for a very long $k$-mers the percolation
may be lost in close analogy with similar behavior observed for $k \times k$ squares~\cite{Nakamura1987}.

\section{\label{sec:concl}Conclusion}

In this paper, the percolation behavior of partially ordered linear $k$-mers on torus  (square lattice with periodic boundary conditions) has been investigated by computer simulations. The length of a $k$-mer varies from $1$ to $512$ and different lattice sizes up to $L = 1024k$ are used. The relaxation random sequential adsorption model~\cite{Lebovka2011PRE} has been used to place the $k$-mers on a lattice. The alignment degree is characterized by order parameter $s=0\dots1$: $s=0$ for isotropic system and $s=1$ for perfectly aligned system.
The behavior of percolation cumulant at the intersection point $R^*$ has been studied in details in dependence on $k$, $s$ and $L$. For isotropic problem the value of position of intersection point remained unchanged within precision of estimation, being $R^*\simeq 0.90$ for all studied length of $k$-mers. The universality of intersection points $R^*$ (i.e., independence of $R^*$ on $k$) has been observed only for isotropic systems, $s=0$.
This universality suggests that $R^*$ can be derived from the work~\cite{Pinson1994JSPh} not only for topological percolation but also for physical one. For anisotropic systems this universality is violated and the value of $R^*$ is dependent upon $k$ and $s$. One can suppose that this violation can reflect the effect of the system anisotropy.

The increase of system ordering always results in increase of percolation threshold $p_c$. The dependencies of  $p_c(k)$ for completely ordered ($s=1$) and partially ordered ($s<1$) systems are obviously different.  For completely ordered systems the percolation threshold $p_c$ monotonically decreased as $k$ increased. The power law relation $p_c\propto 1/k^{\alpha_1}$ ($\alpha_1=0.72 \pm 0.04$) probably reflects effects of the lattice discreteness. For partially ordered systems  the percolation threshold $p_c$ is always a nonmonotonic function of $k$ and for a certain length of $k$-mers $k=k_m$ a minimum of $p_c$ has been observed. It has been assumed that this behavior may reflect the competition of the lattice discreteness (that is dominant at small values of $k$) and the tendency of $k$-mers for stacking, or formation of squarelike blocks (that is dominant at large values of $k$). For completely ordered systems ($s=1$) the ratio of percolation and jamming concentration $p_c/p_j$ is practically constant ($p_c/p_j\simeq 0.715$, at $k>8$). This behavior evidently reflects the presence of some universal connection in the geometry of percolation and jamming~\cite{Vandewalle2000}. For isotropic systems ($s=0$) this ratio is not constant and increased proportionally to $\log_{10} k$.

Our simulations suggest that for $s=0$ the percolation may be lost at $k\gtrapprox 1.2\times10^4$. Additional investigation of percolation with extreme long objects should be performed in future to confirm or reject this prediction.

\begin{acknowledgments}
Work is supported by the Ministry of Education and Science of Russia, project no 1.588.2011.
\end{acknowledgments}

\bibliography{bibfile}

\end{document}